\documentclass[twocolumn,prl]{revtex4}
\usepackage{graphicx}
\newcommand{\ket}[1]{|{#1}\rangle}			
\newcommand{\bra}[1]{\langle{#1}|}

\begin{document}
\title{Unconditionally secure quantum key-distribution with relatively strong signal pulse}

\author{Kiyoshi Tamaki}
\email{tamaki@will.brl.ntt.co.jp}
\affiliation{NTT Basic Research Laboratories, NTT Corporation, 3-1, Morinosato Wakamiya Atsugi-Shi, Kanagawa, 243-0198, Japan\\ CREST, JST Agency, 4-1-8 Honcho, Kawaguchi, Saitama, 332-0012, Japan\\}

\begin{abstract}
We propose an unconditionally secure quantum key distribution (QKD) protocol, which uses a relatively strong signal pulse. While our protocol shares similar security bases as the Bennett 1992 protocol with a strong reference pulse (B92), our scheme uses a smaller number of detectors and it is robust against Rayleigh scattering in an optical fibre. We derive a lower bound of secret key generation rate of our protocol and show that our protocol can cover relatively long distances, assuming precise phase modulations and stable interferometers.

\end{abstract}



\maketitle

\begin{figure}[tbp]
\begin{center}
\includegraphics[scale=0.32]{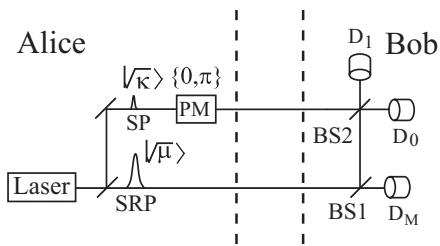}
\end{center} 
\caption{The essential experimental setup for B92. PM: phase modulator. 
\label{fig1}}\end{figure}

Quantum key distribution (QKD) provides a way to share a secret key with arbitrary small leakage of its information to an unauthorized party (Eve). The first QKD protocol, BB84, was proposed in \cite{BB84}, and it originally assumes the use of a single-photon source. In \cite{W03}, it was found that even if we use attenuated laser light, we can still cover long distances with the help of the decoy state method. In this method, the sender (Alice) emits a signal pulse (SP) together with extra pulses (decoy states) whose properties are the same as those of the SP except for their intensities. With the decoy states, Alice and the receiver (Bob) can monitor Eve's action tightly so that they can achieve long distances. One of the drawbacks of this method is the increase of the number of classical communications needed, and it has been reported that the fluctuations of intensities of decoy states decrease the achievable distance significantly \cite{hayashi}. This means that a simpler protocol without decoy states might be preferable in some scenario.

The Bennett 1992 protocol with a strong reference pulse (B92) \cite{B92} uses another approach to cover long distances. In B92, a dim SP is sent together with strong reference pulse (SRP), and the unconditional security of this protocol was proven in \cite{K04, TLKB06}. Assuming typical experimental parameters without taking into account Rayleigh scattering in an optical fibre, in \cite{TLKB06} it is concluded that if the intensity of SP is about $0.1$ and the one for SRP is more than $10^{6.5}$, then long distances can be achieved. However, this strong intensity of the reference light highly causes Rayleigh scattering, and as a result, the achievable distance of practical B92 is very limited due to the bit errors induced by the scattering.

In this paper, we propose an unconditionally secure QKD protocol without decoy states, which shares similar security bases as the B92 and uses the same intensities both for the SP and reference pulse (RP). The intensity can be set to be much weaker than the one for the SRP in the B92. Thus, it is expected that our protocol is free from the Rayleigh scattering problem. A complete security proof for our protocol in terms of achievable distances is still missing. However, we show that even with an unconditional security proof that does not fully capture the security bases, our protocol still can cover relatively long distances assuming precise and stable phase modulations. We expect therefore that our protocol keeps the door open for expanding the achievable distances. Thus, our protocol is not only interesting from a practical point of view, but also it poses an interesting theoretical problem. Note that a similar protocol with strong signal light, homodyne detection, and threshold values has been proposed by Inoue and Hayashi \cite{inoue}.

In this paper, we first explain how the B92 works to illustrate its security essence, and then we introduce a new protocol. Next, we prove unconditional security of the new protocol with additional assumptions on Bob's detectors, and we show some examples of its key generation rate in terms of distances.

In the experiment for the B92, we use double Mach-Zehnder interferometers, however, the essence can be explained by just a single Mach-Zehnder interferometer (see Fig.~\ref{fig1}). In this protocol, Alice prepares a coherent light pulse in a state $\ket{(-1)^{j_{A}}\sqrt{\kappa}}_{\rm SP}\ket{\sqrt{\mu}}_{\rm SRP}$ depending on a random bit value $j_A=0,1$, where $\kappa$ and $\mu$ represent respectively the mean photon number of the SP and the SRP. On the receiving side, Bob uses an asymmetric beam splitter BS1 with reflectivity $\kappa/\mu$, which splits the reference pulse into a weak pulse and a strong one. The intensity of the weak pulse is equalized to that of the incoming signal pulse so that we have an interference in the symmetric beam splitter BS2, i.e., if a ``signal reading detector'' ${\rm D}_0$ $({\rm D}_1)$ clicks, then Bob can infer that Alice has set $j_A=0$ ($j_A=1$). Note that since the incoming SP is very weak, in most cases Bob has no click at the signal reading detectors, which we call inconclusive events. In addition to using the signal reading detectors, Bob needs to test whether the reference pulse always arrives by using a ``monitoring detector'' ${\rm D}_{\rm M}$ just after the BS1. After these measurements are performed, Bob needs to tell Alice whether he has obtained a conclusive result and the monitoring detector has clicked or not.

The essence of the security of the B92 can be captured by considering the following two specific attacks. The first one is so-called beam-splitting attack \cite{DLM06} where Eve uses beam splitters to split some portions of both the signal and the reference pulse to her side. In the B92 the overlap of the two signal states sent by Alice is very large, which means that the probability that both of Eve and Bob have the conclusive results is very small. As a result, Eve's knowledge on Bob's bit values is strongly limited so that a beam-splitting attack is not critical in the B92. 

The second strategy for Eve is an unambiguous state discrimination (USD) attack \cite{DLM06}, where she performs an USD measurement on each signal states sent by Alice. USD succeeds with small probability, and when it dose, Eve can obtain full information on the bit value without introducing any bit error. If it fails, Eve may send a vacuum as a fake signal to Bob, which disguises for signal loss events. If this vacuum induces no bit errors, which is the case for the single-photon B92 (S-B92) \cite{TL04}, then the achievable distances for the B92 is highly limited. Note, however, that Eve needs also to send the SRP in the B92 since Bob tests the its presence with ${\rm D}_{\rm M}$. Thus, sending the vacuum as the fake signal results in a random click because of the SRP, which reveals Eve's existence. In other words, the monitoring detector keeps Eve from performing the USD attack. 

\begin{figure}[tbp]
\begin{center}
\includegraphics[scale=0.33]{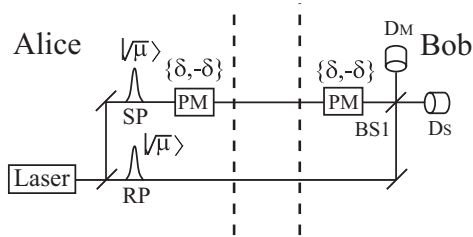}
\end{center} 
\caption{The essential experimental setup for our new protocol.
\label{fig2}}\end{figure}

We call the first basis of the security of the B92 as {\it high nonorthogonality} and the second one as {\it high monitoring ability}. In what follows, we show that these bases can be accomplished by a simple new protocol. Fig.~\ref{fig2} is a schematic of an experimental setup for our new protocol. Here, Bob's interferometer is set up in such a way that if the two incoming pulses have the same phase, then a ``signal reading detector'' ${\rm D}_{\rm S}$ always has a vacuum state. 

Next, we describe steps of our new protocol. (S1) Alice prepares coherent lights pulse in a state $\ket{e^{i (-1)^{j_{A}}\delta}\sqrt{\mu}}_{\rm SP}\ket{\sqrt{\mu}}_{\rm RP}$ according to the random bit value $j_A$, and sends these pulses to Bob. Here, $\delta$ is a small positive number. (S2) Bob randomly chooses his bit value $j_B$ and performs a phase modulation of $e^{i(-1)^{j_{B}}\delta}$ to the incoming signal pulse. (S3) Bob records whether the monitoring detector ${\rm D}_{\rm M}$ clicks and whether the signal reading detector ${\rm D}_{\rm S}$ clicks. (S4) Bob tells Alice whether he has obtained a conclusive event, i.e., ${\rm D}_{\rm S}$ clicks or not. (S5) If Bob has obtained the conclusive event, Alice keeps the corresponding bit value $j_A$. Otherwise, she discards it. (S6) Alice and Bob repeat (S1)-(S5) many times. (S7) If the ratio of the click events of ${\rm D}_{\rm M}$ is low, then they abort the protocol. (S8) Alice and Bob estimate the bit error rate from test bits. Then, they perform classical bit error correction (CEC) \cite{nielsen} and classical privacy amplification (CPA) \cite{nielsen} based on the estimated bit error rate and other available observables, such as the ratio of click events by ${\rm D}_{\rm M}$ and ${\rm D}_{\rm S}$ so that they share a secret key.

Based on the same reason in the B92, this protocol is supposed to be strong against beam-splitting attack and USD attack assuming relatively large $\mu$ and small enough $\delta$. It is not difficult to see in noise-free cases that Alice and Bob share a bit value $j_A=j_B$ with probability $1-e^{-2\eta\mu\sin^2\delta}$, where $\eta$ is a single-photon transmission rate of the channel together with the efficiency of a detector. One can also see that the monitoring detector clicks with probability of $1-e^{-2\eta\mu\cos^2\delta}$. Thus, in order for our protocol to meet {\it high monitoring ability}, $1-e^{-2\eta\mu\cos^2\delta}$ has to be close to $1$. On the other hand, {\it high nonorthogonality} requires that the inner product of Alice's two input state is large, i.e., $e^{-4\mu\sin^2\delta}\sim1$. Combining them together, we conclude that as long as we set $\mu\ge 1/\eta$ and $\delta\le\sqrt{\eta}$, our protocol should be secure based on the same bases as the B92. It follows that when the communication distance $l$ is about 100 (km), we should set $\mu\ge10^{4}$ and $\delta\le10^{-2}$ according to experimental parameters \cite{GYS04} $\eta=0.045*10^{-\frac{0.21 l}{10}}$. This means that our protocol is more robust against Rayleigh scattering than the B92 where we use more than $10^{6.5}$ mean photon number for the SRP. Thus, we expect that our protocol can cover long distances even in practice.

Although we miss a security proof that fully incorporates the {\it high monitoring ability} in our new protocol, we can still prove the unconditional security of our protocol by directly applying the security proof of the S-B92 \cite{TL04}. For the proof, we additionally assume that Bob's detector can discriminate among vacuum, single-photon, and multi-photon, which is helpful to define Bob's qubit space as we will see later. We accordingly re-define the conclusive (inconclusive) event as the case where ${\rm D}_{\rm S}$ (${\rm D}_{\rm M}$) detects a single-photon and ${\rm D}_{\rm M}$ (${\rm D}_{\rm S}$) has a vacuum, and we define all the other events as losses that will be discarded. Following these changes, the modified version of our protocol works in the same manner as the original one. Throughout the proof, we also assume that all imperfections in Alice's and Bob's devices are under Eve's control. For the later convenience, we define $\ket{0}_{\rm SP}\ket{s}_{\rm RP}\equiv\ket{0_z}$ and $\ket{s}_{\rm SP}\ket{0}_{\rm RP}\equiv\ket{1_z}$ as a basis (Z-basis) of a qubit state, where $\ket{s}$ is a single-photon state, and we also define X-basis eigenstate and Y-basis eigenstate as $\ket{j_x}=\frac{1}{\sqrt{2}}(\ket{0_z}+(-1)^j\ket{1_z})$ and $\ket{j_y}=\frac{1}{\sqrt{2}}(\ket{0_z}+i(-1)^j\ket{1_z})$, respectively.

In order to prove the security of our protocol with the additional assumption on Bob's detectors, we convert our protocol into an Entanglement Distillation Protocol (EDP) \cite{SP00}. First, we consider Alice who prepares two qubit in a state $\ket{\Phi}\equiv\frac{1}{\sqrt{2}}(\ket{0_y}_{\rm A}\ket{e^{i \delta}\sqrt{\mu}}_{\rm SP}+\ket{1_y}_{\rm A}\ket{e^{-i\delta}\sqrt{\mu}}_{\rm SP})\ket{\sqrt{\mu}}_{\rm RP}$, sends out the system RP and SP to Bob, and performs Y-basis measurement on the system A. As a result of this, Alice sends $\ket{e^{i \delta}\sqrt{\mu}}_{\rm SP}\ket{\sqrt{\mu}}_{\rm RP}$ and $\ket{e^{-i \delta}\sqrt{\mu}}_{\rm SP}\ket{\sqrt{\mu}}_{\rm RP}$ randomly, which is equivalent to Alice in the actual protocol. This ends the conversion of Alice's side.

As for Bob's part, first note that Bob uses a phase modulator, linear optics, and photon counters. Thus, without changing any measurement outcome, we can assume that Bob's measurement is preceded by a measurement ${\rm Q}$ that measures the total photon number of the incoming signal and reference pulses. Note that Bob has an access to the measurement outcome of ${\rm Q}$ thanks to the additional assumption on his detectors. One can see that Bob's measurement in the subspace containing a single-photon is represented by POVM \cite{nielsen} $\{F_0, F_1, F_{\rm inconc}\}$, where $F_0=(1/2)\hat P(\ket{\overline{\varphi}_{1}})$ ($\hat P(\ket{\psi})\equiv\ket{\psi}\bra{\psi}$), $F_1=(1/2)\hat P(\ket{\overline{\varphi}_0})$, and $F_{\rm inconc}=1-F_{0}-F_{1}$. Here, $F_0$ and $F_1$ correspond to the conclusive event, we define $\ket{\varphi_{j}}\equiv\cos(\delta/2)\ket{0_x}-i(-1)^j\sin(\delta/2)\ket{1_x}$, and $\ket{\overline{\varphi_{j}}}$ is a qubit state orthogonal to $\ket{\varphi_{j}}$. 

A crucial point in the conversion to EDP on Bob's side is that the measurement on a qubit (a single-photon state) can equivalently be executed by applying a filtering operation \cite{G96} whose successful operation is represented by a Kraus operator \cite{nielsen} $F_{\rm s}=\sin(\delta/2)\hat P(\ket{0_x})+\cos(\delta/2)\hat P(\ket{1_x})$, and then performing Y-basis measurement. This equivalence can be seen by noticing that $F_j=\hat P(F_{\rm s}^{\dagger}\ket{j_y})$ and $F_{\rm inconc}=1-F_{\rm s}^{\dagger}F_{\rm s}$. Note that the successful filtering operation corresponds to the conclusive evens. It is not difficult to check in noise and loss free cases that if Bob's filter succeeds, then Alice and Bob share a maximally entangled state (MES) $\frac{1}{\sqrt{2}}(\ket{0_y}_A\ket{0_y}_B+\ket{1_y}_A\ket{1_y}_B)$. Thus, bit values extracted by Y-basis measurement on MES by Alice and Bob are identical and secure since MES is a pure state. 

Note that since Bob's measurement is an USD measurement, our measurement is identical to the one in the S-B92 where nonorthogonal single-photon polarization states are unambiguously discriminated. Actually, POVM in the S-B92 can be immediately obtained by changing $\sin(\delta/2)\rightarrow\alpha$ and $\cos(\delta/2)\rightarrow\sqrt{1-\alpha^2}$ where $\alpha$ characterizes the nonorthogonality of two single-photon polarization states \cite{TL04}. Moreover, mathematical expressions of Alice's nonorthogonal states in our protocol and those in the S-B92 are the same. Thus, there is one-to-one correspondence between our protocol and the S-B92.

In the presence of Eve's intervention, Alice and Bob do not share a MES even if Bob's filter succeeds. A basic idea for proving the security under Eve's intervention is to consider the distillation of a MES from a mixed state. In \cite{SP00}, Shor and Preskill showed that if Alice and Bob can estimate the number of bit errors ($n\Lambda_{\rm bit}$) and phase errors ($n\Lambda_{\rm ph}$) on $n\Lambda_{\rm fil}$ qubit pairs that have passed the filter, then they can distill at least $n\Lambda_{\rm fil}[1-h(\Lambda_{\rm bit}/\Lambda_{\rm fil})-h(\Lambda_{\rm ph}/\Lambda_{\rm fil})]$ ($n\rightarrow\infty$) of MES. Here, $h(x)\equiv-x\log_2x-(1-x)\log_2(1-x)$, and the bit (phase) error represents the case where Alice's and Bob's measurement outcomes differ in Y (X)-basis. Moreover, according to Shor and Preskill's argument, our protocol followed by the EDP and Y-basis measurement is equivalent to our protocol followed by CEC and CPA. Since the bit error rate can be estimated by test bits, if we can estimate the phase error rate, then the security proof ends.

In \cite{TL04}, it is shown that we can estimate the upper bound of the phase error rate for the S-B92 since the filtering operation relates the phase errors with other observables such as bit errors and conclusive events. Thanks to the one-to-one correspondence between our protocol and the S-B92, we are allowed to directly apply this phase error estimation to our protocol so that we obtain the upper bound of phase error rate $\overline{\Lambda_{\rm ph}}$ by solving the following inequality \cite{TL04}
\begin{equation}
 \Lambda_{{\rm fil}}-2\Lambda_{{\rm bit}}\le \sin(\delta) g(C{\bf{z}})\,,
\label{sfinal}
\end{equation}
where $g((a,b,c,d)^T)\equiv \sqrt{ab}+\sqrt{cd}$. In this inequality, ${\bf{z}}\equiv (\Lambda_{\rm s}, \Lambda_{1x}-p_1(1-\Lambda_{\rm s}), \Lambda_{\rm fil}, \Lambda_{{\rm ph}})^T$, where $\Lambda_{\rm s}$ is fraction that Bob obtains the qubit state, $\Lambda_{1x}$ is a probability of Alice having 1 in her Gedanken X-basis measurement, $p_1$ takes values inside $[0,1]$, which must be optimized in such a way that it maximizes $\Lambda_{{\rm ph}}$. 
Finally, $C$ is a matrix whose inverse is expressed as
\begin{equation}
C^{-1}=\left(
\begin{array}{cccc}
1&1 &1 &1 \\
0&0 &1 &1 \\
\sin^2(\delta/2)&\cos^2(\delta/2) &\sin^2(\delta/2) &\cos^2(\delta/2) \\
0&\cos^2(\delta/2) &\sin^2(\delta/2) & 0
\end{array}
\right)\,.
\end{equation}

To illustrate the key generation rate, we consider a channel that maps $\hat P(\ket{e^{\pm i\delta}\sqrt{\mu}}_{\rm SP}\ket{\sqrt{\mu}}_{\rm RP})$ into $(1-p)\hat P(\ket{e^{\pm i\delta}\sqrt{\eta\mu}}_{\rm SP}\ket{\sqrt{\eta\mu}}_{\rm RP})+p\hat P(\ket{s}_{\rm SP}\ket{0}_{\rm RP})$, where $0\le p\le1$. The first part represents losses in a quantum channel, and the second part models dark counts in Bob's detector since it causes a random click. According to the experiment in \cite{GYS04}, $p=1.7\times 10^{-6}$ and $\eta=0.045*10^{-\frac{0.21 l}{10}}$ neglecting alignment errors. Thus, we have $p_{S}=(1-p)2\eta\mu e^{-2\eta\mu}+p$, $\Lambda_{\rm fil}=(1-p)\eta\mu\sin^2(\delta) e^{-2\eta\mu}+p/2$, and $\Lambda_{{\rm bit}}=p/4$.

\begin{figure}[tbp]
\begin{center}
\includegraphics[scale=0.40]{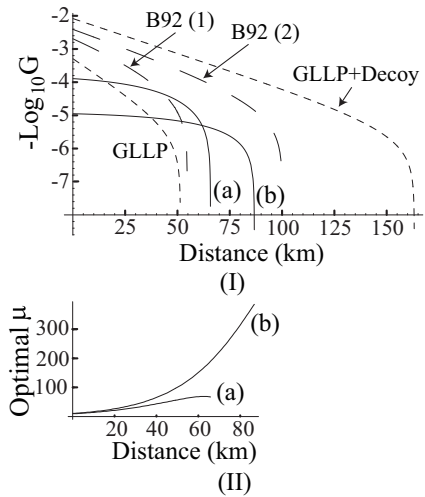}
\end{center}
 \caption{(I) The key generation rates. If Rayleigh scattering is taken into account, the achievable distances of the B92 (B92 (1) and B92 (2)) significantly decrease. (II) The optimal intensity of the light for the modified version of our protocol.
\label{key}}
\end{figure}

In Fig.~\ref{key} (I), we plot the key generation rate $G=\Lambda_{\rm fil}(1-h(\Lambda_{\rm bit}/\Lambda_{\rm fil})-h(\overline{\Lambda_{\rm ph}}/\Lambda_{\rm fil}))$ as a function of the distance (km) between Alice and Bob. In solid lines (a) and (b), we respectively set the precision of the phase modulator as $\Delta\equiv2\delta=\pi/50$ and $\Delta=\pi/150$, and we also plot an optimal $\mu$ to maximize $G$ in Fig.~\ref{key} (II). It is seen that the achievable distance for (a) is $l=66$ (km) and the one for (b) is $l=87$ (km), either of which is larger than the one for BB84 based on the GLLP formula \cite{GLLP02} (dotted line, $l=51$). We have confirmed that 87 (km) is the maximum distances among examples we have tried.

For comparisons, we also plot in Fig.~\ref{key} (I) the key rate for BB84 with infinite number of decoy states \cite{W03} (dotted line, $l=163$), and that for the B92 (dashed line). For the B92, we define a parameter set $(\mu, \kappa, a)$, where $\nu_i=\eta\mu-a\sqrt{\eta\mu}$, and $\nu_f=\eta\mu+a\sqrt{\eta\mu}$ express the photon number regime $[\nu_i, \nu_f-1]$ that the monitoring detector $D_{\rm M}$ has to discriminate from the other regime \cite{TLKB06}. We fix $\kappa=10^{-0.92}$ and $a=3.2$, and we set $\mu=10^5$ for the B92 (1) ($l=55$) and $\mu=10^{6.59}$ for the B92 (2) $(l=100)$. Thus, in the B92 the photon number detected by $D_{\rm M}$ has a crucial role in the achievable distances. We have confirmed in our protocol (in lines (a) and (b)) that Bob's $D_{\rm M}$ fails to click, i.e., fails to detect a single-photon, at least more than 64\% of the instances, which are regarded as the losses, while the loss events in the B92 ($D_{\rm M}$ fails to detect photons inside the photon number regime $[\nu_i, \nu_f-1]$) are negligible \cite{TLKB06}. It follows that our security proof fails to make use of the {\it high monitoring ability}, and we expect that the achievable distances can be expanded by using a better security proof that fully captures this property. Note that if one takes into account Rayleigh scattering, the achievable distances for the B92 significantly decreases. Moreover, the use of a single signal state might be an advantage of our protocol over the decoy state method whose fluctuations make the achievable distance short \cite{hayashi}. 

In summary, we proposed a QKD protocol that shares similar security bases as the B92 and might be robust against the Rayleigh scattering. We have shown that even with an unconditional security proof that does not fully capture the security bases, our protocol still can cover relatively long distances assuming precise phase modulations and stable interferometer.  We leave a security proof that fully captures the security bases of our protocol for the future works.

Helpful discussions with colleagues, including T. Honjo, H. Takesue, T. Yamamoto, M. Koashi, especially K. Inoue, T. Tsurumaru, and M. Curty are gratefully acknowledged. We thank NICT for the support.

\end{document}